\documentclass[aps,prd,twocolumn,showpacs]{revtex4}
\usepackage{amssymb,amsmath,bm,graphicx,thmsa}
\begin{document}

\title{Well-posed first-order reduction of the characteristic problem of the
linearized Einstein equations}

\author{Simonetta Frittelli}
\email[]{simo@mayu.physics.duq.edu}
\affiliation{Department of Physics, Duquesne University,
        Pittsburgh, PA 15282}
\affiliation{Department of Physics and Astronomy, University of Pittsburgh,
       Pittsburgh, PA 15260}

\date{\today}

\begin{abstract}

A choice of first-order variables for the characteristic problem of the
linearized Einstein equations is found which casts the system into manifestly
well-posed form. The concept of well-posedness for characteristic problems
invoked is that there exists an \textit{a priori} estimate of the solution of
the characteristic problem in terms of the data. The notion of manifest
well-posedness consists of an algebraic criterion sufficient for the existence
of the estimates, and is to characteristic problems as symmetric hyperbolicity
is to Cauchy problems.  Both notions have been made precise elsewere. 

\end{abstract}
\maketitle

\section{Introduction} \label{sec:1}


Well-posedness is a concept naturally associated with initial-value problems,
whereby the norm of the solution at some time in the evolution is bounded by
the norm of the initial data, independently of the data. The norm of the
solution is defined by integrating on a time-slice a positive-definite
quadratic expression in the fields. The norm of the data is the same
positive-definite quadratic expression integrated on the initial slice. A
well-posed initial-value problem turns out solutions that are stable with
respect to small variations of the initial data, and is therefore naturally
interpreted as a necessary condition for the existence of a stable numerical
scheme (of the finite-difference kind) for the problem \cite{kreissbook}. 

The well-posedness of the initial-value problem of the Einstein equations has
received much attention since the pioneering work by
Choquet-Bruhat~\cite{choquet83}, and has turned out a rich spectrum of
equivalent formulations of the Einstein equations with differing properties,
such as number of variables, number and nature of the associated characteristic
speeds and allowed gauge choices~\cite{lecolivrev}.  A strong motivation for
such attention is the expectation that well-posed formulations might play a
main role in extending the run time of numerical simulations. 

The understanding of the characteristic problem of the Einstein equations seems
to have been evolving in the opposite direction. With no obvious issues of
gauge and with a reduced number of fields, numerical simulations of spacetimes
via the characteristic formulation have been performing at long running times
with comparatively high efficiency ~\cite{longruncharac,mbh}, and the attention
seems to have centered around improved accuracy~\cite{gomez2001}. But, because
little if anything is known about the stability of solutions of characteristic
problems of partial differential equations, interest in the concept of a
well-posed characteristic formulation of the Einstein equations has been rather
limited~\cite{baleanthesis}. 

The main obstacle to the concept of well-posedness of a characteristic problem
seems to be the peculiar split of the data set into a ``normal'' subset which
is usually interpreted as initial data, and a ``null'' subset which has usually
been thought to have no analog in the initial-value framework.  The ``null''
subset has been interpreted sometimes as a flow of information through the
boundary of the region where the solution is being sought.  So if one were to
ask if there exists an \textit{a priori} estimate for the solution of a
characteristic problem, how should the answer even be framed? What is the
information available \textit{a priori} of the solution? The question is
confusing because the null subset of data must be prescribed on a surface that
is transverse to the characteristic initial slice and that contains the
direction of evolution. One must then know such data variables at all evolution
times, therefore it seems hardly appropriate to call it \textit{a priori}
information.  But if the \textit{a priori} information is confined to only the
normal data, then it is impossible to obtain a truly \textit{a priori} estimate
because the solution will be affected by the ``flow of information''.  One must
then include the null data somehow in the estimate, and one way to do so for
the characteristic problem of the wave equation has been proposed by Balean and
Bartnik~\cite{baleanthesis,balean1,balean2}.  

But in essence, \text{a priori} information is any data that are known before
the solution is known. There is really no issue of ``evolution time''. If we
allow ourselves to regard the null subset as a set of \textit{a priori}
information as good as the normal subset, then we have the same number of data
as the initial value problem of the same equations, and the issue is simply to
find a way to compare the size of the solution within the region of interest
with the size of the data on both data surfaces. More precisely, if the $n$
dependent variables of the solution of the problem are represented as a vector
$v$, the $m (<n)$ null data are represented by $w^0$ and the $n-m$ normal data
are represented by $q^0$, then the estimate would read
\[
	||v||^2 \le K (||w^0||^2 + ||q^0||^2)
\]

\noindent where $K$ would be a number that would not depend on $(w^0,q^0)$. The
norms would have to be defined appropriately in each space. Clearly an
inequality of this kind would ensure that vanishing values of $(w^0,q^0)$ on
their respective data surfaces would return a vanishing solution $v$, and the
size of small perturbations of the data $(w^0,q^0)$ on their respective data
surfaces would control the variation of the solution. The term  $||w^0||^2$
would be the integral of a quadratic expression in the null data on the
corresponding data surface.  Whether we refer to it as data or as a flow of
energy is quite immaterial, but the parallel with initial value problems is
maintained if we choose to call it \textit{data}.  

With this viewpoint, a notion of well-posedness for generic linear homogeneous
characteristic problems can be formulated in reference to the problem's
almost-canonical form\cite{lincharac}.  Even though the choice of coordinates
is not relevant to the essence of a characteristic problem, let's assume that
the equations are written in terms of coordinates adapted to the characteristic
surfaces, so we have $(u,x,x^i)$ with $i=1\ldots r-2$ and with $u$ labeling the
characteristic surfaces.  Then one can always find a choice of dependent
variables $v=(w,q)$ with $w=w_1,\ldots,w_m$ and $q=q_1,\ldots,q_{n-m}$ so that
the system of equations takes the form    
\begin{subequations}
\begin{eqnarray} 
N^u q,_u + N^x q,_x + N^i v,_i + N^0 v &=& 0	\\ 
               w,_x + L^i v,_i + L^0 v &=& 0 
\end{eqnarray}
\end{subequations}

\noindent where the $n-m$ square matrix $N^u$ is non-degenerate.  We refer to
this as the almost-canonical form of the characteristic problem (the canonical
form has the identity matrix in the place of $N^u$ and is achieved by the
transformation $q\to N^u q$, which always exists because $N^u$ is
non-degenerate). This system of equations requires the values of $q$ to be
prescribed at $u=0$ and the values of $w$ to be prescribed at $x=0$ in order
for a unique solution to exist in the region to the future of $u=0$ and to the
right of $x=0$, that is: a total of $n$ functions of $r-1$ variables
~\cite{duff}.  We will say that the problem is \textit{well posed} if there
exists a number $\kappa$ independent of $(q^0,w^0)$ and a positive-definite 
$n-$dimensional symmetric matrix $H$ such that 
\begin{equation}\label{eq:genestimate}
	||v||^2_T \le e^{\kappa T} (||w^0||^2 + ||q^0||^2)
\end{equation}

\noindent where 
\begin{subequations}
\begin{eqnarray}
	||v||_T^2 &\equiv& 
	\int_{u+x=T} \hspace{-0.4cm} (v,Hv) \\
	||q^0||^2 &\equiv& \int_{u=0} (q^0,q^0)  \\
	 ||w^0||^2&\equiv& \int_{x=0} (w^0,w^0)
\end{eqnarray}
\end{subequations}

\noindent In each case, the notation $(\, , \, )$ denotes the standard
Euclidean scalar product of the appropriate dimension; e.g., $(w,w)\equiv
\sum_{I=1}^m w_I^2$.  The notion of well-posedness thus consists of the norm of the
solution with respect to a spacelike surface $u+x = T$ being bounded by the
norm of the data, irrespectively of the data. The surface $u+x = T$ intersects
the data surfaces $u=0$ and $x=0$, and the fixed value of $T$ is arbitrary, but
finite, and may be not expected to be large if the problem has non-constant
coefficients and non-vanishing undifferentiated terms. 

An equivalent representation of an almost-canonical characteristic problem
is the form
\begin{equation}
	C^a v,_a + D v =0,
\end{equation}

\noindent where $a=u,x,i$, the principal matrix $C^u$ has a block diagonal form
diag$(N^u,0)$, the principal matrix $C^x$ has a block diagonal form
diag$(N^x,1)$, and the remaining principal matrices $C^i$ are completely
arbitrary. We say that the  characteristic problem is \textit{manifestly well
posed} if the following two conditions are satisfied: \newline 
\textbf{ i)} All the principal matrices $C^a$ are symmetric \newline
\textbf{ ii)} The normal block of the principal-$u$ matrix, $N^u$, is
positive-definite and the normal block of the principal-$x$
matrix, $N^x$, is non-positive definite but such that $N^u+N^x$ is positive
definite (that is,  $-N^u < N^x \le 0$). 

If these conditions are satisfied, then one can take $H \equiv C^u+C^x$ and
there exists $\kappa$ such that (\ref{eq:genestimate}) holds. The proof of this
statement is developed elsewhere~\cite{lincharac}, and parallels an argument
first used to determine \textit{a priori} estimates for the characteristic
problem of the wave equation \cite{cmpwave}. The notion of manifest
well-posedness of characteristic problems as defined here is a close analog of
the notion of symmetric hyperbolicity for initial-value
problems~\cite{kreissbook}. The conditions \textbf{ (i)} and \textbf{(ii)}
together constitute a sufficient algebraic criterion for well-posedness of
characteristic problems. Clearly, the criterion is not a necessary condition
(just as symmetric hyperbolicity is not necessary for well-posedness). 

In Section \ref{sec:2} we introduce the characteristic problem of the Einstein
equations in second-order form --the form in which the equations are most
commonly used for numerical implementation-- and linearize them around
Schwarzschild spacetime. Subsequently, we find a set of first-order variables
that cast the characteristic problem into manifestly well-posed form in the
sense introduced in the previos paragraph. We conclude that the linearized
characteristic problem of the Einstein equations admits a well-posed
formulation. Section \ref{sec:3} offers concluding remarks regarding the reach
of the result and some remaining open questions.

\section{The characteristic problem in general relativity}\label{sec:2}

\subsection{Original problem}

The line element $ds^2=g_{uu}du^2 + g_{ur}dudr +g_{ua}dudx^a+g_{ab}dx^adx^b$ in coordinates adapted to a slicing by null hypersurfaces 
labeled by a coordinate $u$ is
\begin{eqnarray}
	ds^2
   &=&
	-\left(\frac{e^{2\beta}V }{r} 
		-r^2 h_{ab}U^aU^b\right) du^2
	-2e^{2\beta} dudr \nonumber\\
&&
	-2r^2h_{ab}U^b dudx^a
	+r^2h_{ab}dx^adx^b,
\end{eqnarray}

\noindent where $x^a=(\theta,\phi)$ label null geodesics in the constant-$u$
slice, and $\det(h_{ab})=\sin^2\theta$, so that $r$ is the luminosity distance
and functions as a parameter along the geodesics.  We define $h^{ab}$ as the
2-inverse of $h_{ab}$, via $h^{ab}h_{bc}=\delta^a{}_c$. Enough coordinate
conditions have thus been imposed to leave only six metric functions to
determine by means of the Einstein equations. 

Sachs' approach \cite{sachs} to the Einstein equations adapted to a null
slicing considers the components of the Einstein equations $G_{\mu\nu}=0$ in a
frame adapted to the null slicing. Such a frame has the direction $\ell^\alpha$
normal to the slicing as its main leg. With the current signature convention,
$(-,+,+,+)$, the second leg of the frame is another null vector $n^\alpha$ such
that $\ell^\alpha n_\alpha=-1$. The frame is completed by a complex null vector
$m^\alpha$ normalized by $m^\alpha\bar{m}_\alpha=1$ that is perpendicular to
both $\ell^\alpha$ and $n_\alpha$.  The complex vector $m^\alpha$ is equivalent
to two real orthonormal spacelike vectors $e_1^\alpha,e_2^\alpha$ tangent to
the surface of fixed value of $u$, but the complex combination $m^\alpha =
(e_1^\alpha+ie_2^\alpha)/\sqrt{2}$ is usually preferred because it facilitates
some calculations while leading to more compact expressions. Up to scaling, the
normal direction is $\ell^\alpha\equiv g^{\alpha\beta} u,_\beta = g^{\alpha u}$,
and is tangent to the null geodesics of fixed value of $x^a$ on each slice,
being thus proportional to $\partial/\partial r$. 

In the current notation, Sachs' approach consists in splitting the Einstein
equations into three sets. The first set contains six \textit{main equations},
which themselves split into the four \textit{hypersurface equations},
$G_{\mu\nu}\ell^\mu\ell^\nu = G_{\mu\nu}\ell^\mu m^\nu =
G_{\mu\nu}m^\mu\bar{m}^\nu=0$, and the two \textit{main equations}, 
$G_{\mu\nu}m^\mu m^\nu =0$.  The second set has only one equation, the
\textit{trivial equation}, $G_{\mu\nu}\ell^\mu n^\nu =0$. The third set
contains the three \textit{supplementary conditions}, $G_{\mu\nu}n^\mu m^\nu = 
G_{\mu\nu}n^\mu n^\nu =0$.  The six main equations are the equations that
determine the six unknown metric functions $h_{ab}, \beta, V, U^a$. Sachs
argues that by virtue of the Bianchi identities ($\nabla_\mu G^{\mu\nu}=0$),
the remaining equations can be thought of as constraints on the data for the
unknown metric functions and can be ignored for our purposes.  

The main equations have been used for numerical simulations for some time,
starting with cases with high symmetry 
\cite{welling,piran,bishopd'inv,stewartchar,clarked'inv,robernullcone} and
moving to full generality by 1996~\cite{cce} (see \cite{jefflivrev} for
citations and an up-to-date review of the numerical implementation of Sachs
characteristic approach to the Einstein equations).  To my knowledge, the
hypersurface equations appeared explicitly for the first time in \cite{jeff83},
taking the following form:
\begin{subequations}
\begin{eqnarray}
	0
   &=&
	 \beta_{,r}
	-\frac{r}{16}h^{ac}h^{bd}h_{ab,r}h_{cd,r}	\\
	0
   &=&
	 (r^4e^{-2\beta}h_{ac}U^c_{,r})_{,r}
	-2r^4\left(\frac{1}{r^2}\beta_{,a}\right)_{,r}
	+r^2h^{cd}D_ch_{ad,r}				\nonumber\\
&& \\
	0
   &=&
	2V_{,r}
	-e^{2\beta}{\cal R}
	+2e^{2\beta}(D^cD_c\beta+D^c\beta D_c\beta)\nonumber\\
&&
	-\frac{1}{r^2}D_c(r^4U^c)_{,r}
	+\frac{r^4e^{-2\beta}}{2}h_{cd}U^c_{,r}U^b_{,r}.					
\end{eqnarray}
\end{subequations}

\noindent Here $D$ is the covariant derivative associated with $h_{ab}$
(i.e., such that $D_ch_{ab}=0$), and $\cal R$ is the curvature scalar of the
2-surfaces given  by the intersection of the constant-$u$ and constant-$r$
hypersurfaces.  The evolution equations as they appeared in \cite{jeff83} are
\begin{eqnarray}\label{pre-evol}
	0 
   &=&
	m^am^b\Big(r(rh_{ab,u})_{,r}
	-\frac12 (rVh_{ab,r})_{,r}		\nonumber\\
&&
	-2e^{2\beta}(D_aD_b\beta+D_a\beta D_b\beta)
	+h_{c(a}D_{b)}(r^2U^c)_{,r}\nonumber\\
   & &
	-\frac{r^4e^{-2\beta}}{2}h_{ac}h_{bd}U^c_{,r}U^d_{,r} 
	 +\frac{r^2}{2}h_{ab,r}D_cU^c\nonumber\\
&&
	+r^2U^cD_ch_{ab,r}			
	-r^2D^cU_{(b}h_{a)c,r}
	+r^2D_{(b}U^ch_{a)c,r}\Big)\nonumber\\
&&								
\end{eqnarray}

\noindent Since we are interested in a real (as opposed to complex) version of
the evolution equations explicitly in terms of the metric variables, rather
than implicitly through the vector $m^a$, we wish to ``un-contract'' the
vectors $m^am^b$.  Given a generic tensor quantity $H_{ab}$, the contraction
$m^am^bH_{ab}=0$ kills both its skew-symmetric part and its trace, so the
``uncontracted'' equivalent of (\ref{pre-evol}) is symmetric in $a,b$ and
traceless:
\begin{eqnarray}
	0 
   &=&
	r(rh_{ab,u})_{,r}
	-\frac12 (rVh_{ab,r})_{,r} \nonumber\\
&&
	-2e^{2\beta}(D_aD_b\beta+D_a\beta D_b\beta)
	+h_{c(a}D_{b)}(r^2U^c)_{,r}\nonumber\\
   & &
	-\frac{r^4e^{-2\beta}}{2}h_{ac}h_{bd}U^c_{,r}U^d_{,r}	
	+\frac{r^2}{2}h_{ab,r}D_cU^c \nonumber\\
&&
	+r^2U^cD_ch_{ab,r}
	-r^2D^cU_{(b}h_{a)c,r}
	+r^2D_{(b}U^ch_{a)c,r}					\nonumber\\
   & &
	+\frac12h_{ab}\bigg(
		r^2h^{cd}{}_{,r}\Big(   h_{cd,u}
					-\frac{V}{2r}h_{cd,r}\Big)\nonumber\\
&&
		+2e^{2\beta}(D^cD_c\beta+D^c\beta D_c\beta)
		-D_c(r^2U^c)_{,r} \nonumber\\
&&
		+\frac{r^4e^{-2\beta}}{2}h_{cd}U^c_{,r}U^d_{,r}
		      \bigg)				\label{mainh}					
\end{eqnarray}

\noindent In the following, we linearize the equations around a Schwarzschild
background.  In our current notation, Schwarzschild spacetime is given by
$U^a=\beta=0$, $V=r-2m$ and $h_{ab}=$diag$(1,\sin^2\theta)\equiv q_{ab}$. For
spacetimes that are close to Schwarzschild we can define first-order variables
$\tilde{h}_{ab}$ and $\tilde{V}$ via 
\begin{eqnarray}
	h_{ab}&=&q_{ab}+\tilde{h}_{ab}\\
	V     &=&r-2m+\tilde{V}.
\end{eqnarray}

\noindent This implies that the covariant derivative $D_a$ is equal to the
covariant derivative on the sphere $\nabla_a$ up to terms of first order. 
Keeping only first-order terms in $\tilde{h}_{ab}, \tilde{V}, \beta$ and
$U^a$ in the equations one obtains their linearized version:
\begin{subequations}
\begin{eqnarray}
	r(r\tilde{h}_{ab})_{,ur} &&\nonumber\\
	-\frac{(r\!-\!2m)}{2}(r\tilde{h}_{ab})_{,rr}
	&=&
	-\frac{2m}{r}((r\tilde{h}_{ab})_{,r}-\tilde{h}_{ab})
	+2\nabla_a\!\nabla_b\beta \nonumber\\
&&
	-q_{ab}\nabla^c\nabla_c\beta
	-q_{c(a}\nabla_{b)}\!(r^2U^c)_{,r}\nonumber\\
&&
	+\frac12q_{ab}\nabla_c(r^2U^c)_{,r}	\label{h,ur}
\end{eqnarray}
\begin{eqnarray}
	\beta_{,r}
	&=&
	0					\label{beta,r}	\\
	r(q_{ac}(r^2U^c)_{,r})_{,r}
	&=&
	2rq_{ac}U^c
	-4\beta_{,a}
	-\nabla^b(r\tilde{h}_{ab})_{,r}	\nonumber\\
&&
	+\nabla^b\tilde{h}_{ab}			\label{U,r}	
\\
	\tilde{V}_{,r}
	&=&
	\frac12 q^{ac}q^{bd}\tilde{h}_{ab,cd}
	-\nabla^c\nabla_c\beta	\nonumber\\
&&
	+\frac12 \nabla_c(r^2U^c)_{,r}
	+r\nabla_cU^c
	+2\beta	\nonumber\\
&&
	+\frac{3\cos\theta}{2\sin\theta}\tilde{h}_{\theta\theta,\theta}
	-\tilde{h}_{\theta\theta}.		\label{V,r}
\end{eqnarray}	
\end{subequations}

\noindent In the process to obtain (\ref{V,r}) we used 
\begin{eqnarray}
{\cal R} 
   &=& 
	2+\frac{2}{\sin^2\theta}\tilde{h}_{\theta\phi,\theta\phi}
	+\tilde{h}_{\theta\theta,\theta\theta}
	+\frac{1}{\sin^4\theta}\tilde{h}_{\phi\phi,\phi\phi}\nonumber\\
&&
	+ \frac{3\cos\theta}{2\sin\theta}\tilde{h}_{\theta\theta,\theta}
	-\tilde{h}_{\theta\theta}
\end{eqnarray}

\noindent together with the fact that 
\begin{equation}
	q^{ab}\tilde{h}_{ab}=0
\end{equation}

\noindent by virtue of det$(h_{ab})=$det$(q_{ab})$. 

\subsection{Well-posed first-order form of the linearized characteristic 
problem}

We define now a set of variables that casts the original system
into first-order form, namely, such that no second-derivatives appear. 
Clearly there is an infinite number of ways to achive this purpose, one of
which has been used previously~\cite{lui}. However,
the following choice is particularly convenient for our current purposes:
\begin{subequations}
\begin{eqnarray}
P_{ab}&\equiv& (r\tilde{h}_{ab})_{,r},	\\
Q_a   &\equiv& q_{ac}(r^2U^c)_{,r} - 2\beta_{,a},	\\
T_a   &\equiv& \beta_{,a},			\\
J_a   &\equiv& \nabla^b\tilde{h}_{ab} + q_{ac}(r^2U^c)_{,r}.			
\end{eqnarray}
\end{subequations}

\noindent It is clear that $\tilde{h}_{ab,r}$ is encoded in $P_{ab}$ and that
the set $(\beta_a, U^a_{,r})$ is encoded in the set $(Q_a,T_a)$, since given
$(Q_a,T_a)$ we can solve uniquely for $(\beta_a, U^a_{,r})$.  In terms of these
variables, equations (\ref{h,ur}), (\ref{U,r}) and (\ref{V,r}) read
\begin{subequations}
\begin{eqnarray}
 	rP_{ab,u}
	-\frac{(r\!-\!2m)}{2}P_{ab,r} &&\nonumber\\
	+\nabla_{(b}Q_{a)}
	-\frac12q_{ab}\nabla^cQ_c		
	&=&
	-\frac{2m}{r}(P_{ab}-\tilde{h}_{ab})	\label{P,ur}\\
	rQ_{a,r}
	+\nabla^bP_{ab}	
	&=&	
	2rq_{ac}U^c
	-4T_a
	+J_a					\label{Q,r}\\
	\tilde{V}_{,r}
	+q^{bd} T_{b,d}-\frac12 q^{bd}J_{b,d} &&\nonumber\\
	-\frac{3\cos\theta}{2\sin\theta}\tilde{h}_{\theta\theta,\theta}
	-r\nabla_cU^c  && \nonumber\\
	\hspace{-1cm}+q^{ac}\Big(q^{bd}_{,c}\tilde{h}_{ab,d}
	- q^{bd}\Gamma_{ad}^e\tilde{h}_{eb,c} &&\nonumber\\
		    -q^{bd}\Gamma_{bd}^e\tilde{h}_{ae,c}
		\Big)	
	&=&
	2\beta
	-\tilde{h}_{\theta\theta} 
	-\frac12 q^{ad}\Gamma_{ad}^cQ_c \nonumber\\
&&
	+q^{ac}\left( q^{bd}\Gamma_{ad}^e\right),_c\tilde{h}_{eb}\nonumber\\
&&
	+q^{ac}\left( q^{bd}\Gamma_{bd}^e\right),_c\tilde{h}_{ae}\label{lastV,r}
\end{eqnarray}

\noindent The quantities $\Gamma^c_{ab}$ that appear in the right-hand side of
(\ref{lastV,r}) are the Christoffel symbols of the metric of the 2-sphere,
$q_{ab}$, thus representing given functions of $\theta$, but not of the
dependent variables.  As it stands, Eq.~(\ref{lastV,r}) is sufficiently
explicit for our purposes, as will be made clear shortly. Additionally, we have
\begin{eqnarray}
	T_{a,r}
	&=& 0			\label{T,r}\\
	rJ_{a,r}
	&=& Q_a -2T_a+2rU_a,	\label{J,r}
\end{eqnarray}

\noindent as a consequence of (\ref{beta,r}) and of (\ref{U,r}).  Furthermore,
by definition, we have the additional equations
\begin{eqnarray}
	r\tilde{h}_{ab,r}	
	&=& P_{ab} -\tilde{h}_{ab} 	\label{h,r}\\
	r^2U^a_{,r}
	&=&
	Q^a + 2T^a -2rU^a.		\label{lastU,r}
\end{eqnarray}
\end{subequations}

\noindent Equations (\ref{beta,r}), (\ref{P,ur}), (\ref{Q,r}), (\ref{lastV,r}),
(\ref{T,r}), (\ref{J,r}), (\ref{h,r}) and (\ref{lastU,r}) constitute a linear
homogeneous characteristic system of equations for the variables
$(\tilde{h}_{ab}, \beta, U^a, \tilde{V}, P_{ab}, Q_a, T_a, J_a)$ in canonical
form.  The two variables $P_{ab}$ are the normal variables of the problem,
whereas the remaining twelve variables are all null variables.  The system has
a unique solution for normal data $P_{ab}^{(0)}\equiv (q_1^0,q_2^0)\equiv q^0$
given on the slice $u=0$, and null data $(\tilde{h}_{ab}^{(0)}, \beta^{(0)},
{U^a}^{(0)},\tilde{V}^{(0)}, Q_a^{(0)}, T_a^{(0)}, J_a^{(0)})\equiv
(w_1^0,\ldots,w_{11}^0) \equiv w^0$ given on the worldtube
$r=r_0$~\cite{duff}. 

For our purposes, however, we point out the following fact, particular to the
linearized regime of the characteristic equations.  The variable $\tilde{V}$
appears \textit{only} in Eq. (\ref{lastV,r}).  The rest of the equations in the
system  yield the values of the remaining variables, which can be taken as a
source for $\tilde{V}$ in Eq. (\ref{lastV,r}). Thus the variable $\tilde{V}$
can be thought of as decoupled from the system, in the linearized regime. In
other words, the system can be solved for $(P_{ab},Q_a,T_a, J_a,\tilde{h}_{ab},
\beta, U^a)$ and then the solutions can be used to integrate $\tilde{V}$ from
equation (\ref{lastV,r}).  The subsystem that yields $(P_{ab},Q_a,T_a,
J_a,\tilde{h}_{ab},\beta, U^a)$ is also a characteristic system in canonical
form, with a two-dimensional normal space, and an eleven-dimensional null space. In
the following, this is the point of view that we take in order to argue that
the linearized system is well posed in the sense of Section~\ref{sec:1}.  

The subsystem obtained by leaving out Eq. (\ref{lastV,r}) has the form 
\begin{equation}
	C^uv_{,u} 
	+C^rv_{,r}
	+C^av_{,a}
	+
	Dv =0
\end{equation}

\noindent where $v=(P_{ab},Q_a,T_a, J_a,\tilde{h}_{ab}, \beta, U^a)$. The
matrix $C^u$ has the form diag$(1,1,0,\ldots,0)$, whereas  the matrix $C^r$ has
the form diag$(-(r-2m)/2r,1,\ldots,1)$.  Additionally, the matrices $C^a$ are
symmetric, coupling only the variables $Q_a$ and $P_{ab}$.
The system is linear and homogeneous and meets the criterion for well posedness
outside of the event horizon, since the normal block  of $C^r$ is negative
definite for all $r>2m$, and the normal block of $C^u+C^r$ is positive definite
for all values of $r$. This implies that there exists a constant $\kappa$
independent of the data $v^0 \equiv(q^0,w^0)$ such that 
$
	||v||_T^2 \le e^{\kappa T}\left(||q^0||^2 
			 	       + ||w^0||^2 \right)
$
where the norms are defined as follows
\begin{subequations}
\begin{eqnarray}
	||v||_T^2 &\equiv& \int_{u+r=T} \hspace{-0.4cm} v(C^u+C^r)v \\
	||q^0||^2 &\equiv& \int_{u=0} {\textstyle \sum} (q_I^0)^2 \\
	 ||w^0||^2&\equiv& \int_{r=r_0} {\textstyle \sum} (w_I^0)^2
\end{eqnarray}
\end{subequations}

\noindent As argued in Section~\ref{sec:1}, this in turn ensures that the
solutions are stable with respect to small perturbations of the data. 

Notice that the surface $u+r=T$ for fixed value of $T$ is a spacelike surface
with respect to the hyperbolic operator because $C^u+C^r$ is positive definite.
Additionally, one can verify that the foliation by constant values of $u+r$ is
also spacelike with respect to the Schwarzschild metric, and thus  it
corresponds to some timelike coordinate for Schwarzschild space other than the
standard Schwarzschild time. This is, of course, immaterial for the purpose of
defining a positive-definite norm of the solution $v$.

\section{Concluding remarks} \label{sec:3}

The main result is a formulation of the linearized characteristic problem of
the Einstein equations that is a counterpart to a symmetric hyperbolic
formulation of the 3+1 Einstein equations with respect to the stability of the
solutions under small variations of the data.  The equations are
(\ref{beta,r}), (\ref{P,ur}), (\ref{Q,r}), (\ref{T,r}), (\ref{J,r}),
(\ref{h,r}) and (\ref{lastU,r}), which are thought to be solved as a
characteristic problem of partial differential equations, from normal data  
$P^0_{ab}(r,x^a)$ on the surface $u=0$ and null data
$(Q_a^0(u,x^a),T_a^0(u,x^a), J_a^0(u,x^a),\tilde{h}_{ab}^0(u,x^a),
\beta^0(u,x^a), U^a_0(u,x^a))$ on the worldtube $r=r_0$ outside of the event
horizon of the background Schwarzschild black hole. The solution to this
problem is used as a known source for the remaining metric variable $\tilde{V}$
which is obtained by integrating an ordinary differential equation in $r$,
Eq.~(\ref{lastV,r}), from data $\tilde{V}^0(u,x^a)$ given on the worldtube.  
The problem has 14 variables in all, a significant reduction with respect to the
standard hyperbolic formulations of the Einstein equations which generically
require 30 dependent variables.  

With respect to the reduction in the number of variables, it is perhaps
appropriate to point out that the reduction has not so much to do with the use
of characteristic coordinates, but much to do with the fact that the original
equations are of second-order. In this respect, consider the original
second-order formulation of the Einstein equations in terms of one timelike and
three spacelike coordinates: they are six second-order equations for six
variables where all second-order derivatives of all variables occur
generically.  The first-order representation of such a system generically
requires that all first-derivatives of the six variables be promoted to new
variables, that is: an addition of $6\times 4$ new variables.  Hence the
generic number of variables of a first-order formulation of the Einstein
equations: 30.  Any one of such formulations is hyperbolic in the sense that it
admits characteristic surfaces: surfaces that are not appropriate as
initial-data surfaces for the Cauchy problem.  One can write the characteristic
problem of such a hyperbolic formulation, but this problem will have no less
than 30 variables -- the characteristic problem of any first-order hyperbolic
system of equations has as many variables (and data) as the initial-value
problem~\cite{duff,lincharac}. 

But now consider writing the second-order form of the Einstein equations in
terms of a coordinate that labels characteristic surfaces of the second-order
system.  By construction, the second-order derivatives with respect to the
characteristic coordinate \textit{do not occur} in the equations (that's why
the coordinate is characteristic). Therefore, the first-order representation of
the characterisic problem of the Einstein equations must have at least six
fewer variables than the generic first-order hyperbolic formulation, that is,
at most 24 variables. A simpler instance of the non-commutativity of the
operations of casting a system into characteristic form and of reducing to
first-order is that of the two characteristic problems of the (scalar) wave
equation. The problem of reducing to first-order and then going to a
characteristic formulation can be found in ~\cite{lincharac}, whereas the
problem of first transforming to characteristic coordinates and then going to
first-order appears, for instance, in ~\cite{cmpwave}.  

The characteristic problem that we have examined here does not return a
solution to the Einstein equations as it stands, because we have neglected four
equations: the ``trivial'' equation and the ``supplementary
conditions''~\cite{sachs}, which do not seem to appear in explicit form in the
literature. Irrespectively of their explicit form, these equations are
essential if one is to obtain a metric that solves the Einstein equations via
the characteristic approach. Nevertheless, considering them jointly with the
main equations for a global statement of well-posedness is outside of the scope
of the current work. 

A number of issues remain open to further study. One question is whether
estimates can be established for the derivatives of the solution of the
characteristic problem of the Einstein equations. As stated in
~\cite{lincharac}, manifest well-posedness of a linear homogeneous
characteristic problem in general does not guarantee the existence of \textit{a
priori} estimates for the derivatives (in stark contrast to the case of
initial-value problems~\cite{courant}). Another open issue is how the
well-posedness of the characteristic problem relates to well-posedness of a
hyperbolic formulation of the Einstein equations.  

Lastly, one may want to consider other first-order reductions of Sachs'
characteristic problem in order to establish estimates that are of greater
relevance to the non-linear case, in particular the 18-variable first-order
reduction developed in \cite{quasicharac}. On the other hand, whether or not
the argument used in the linearized case is relevant to the full non-linear
case is not really an issue at this time, since a concept of well-posedness for
non-linear characteristic problems is not available yet.   


\begin{acknowledgments}
I am deeply indebted to Roberto G\'{o}mez for numerous conversations. This
work was supported by the NSF under grants No. PHY-9803301, No.
PHY-0070624 and No. PHY-0244752 to Duquesne University. 
\end{acknowledgments}


\end{document}